\documentclass[conference]{IEEEtran}
\IEEEoverridecommandlockouts
\usepackage{cite}
\usepackage{amsmath,amssymb,amsfonts}
\usepackage{algorithmic}
\usepackage{graphicx}
\usepackage{textcomp}
\usepackage{xcolor}
\def\BibTeX{{\rm B\kern-.05em{\sc i\kern-.025em b}\kern-.08em
    T\kern-.1667em\lower.7ex\hbox{E}\kern-.125emX}}
\makeatletter
 \let\old@ps@headings\ps@headings
 \let\old@ps@IEEEtitlepagestyle\ps@IEEEtitlepagestyle
 \def\confheader#1{%
 \def\ps@IEEEtitlepagestyle{%
 \old@ps@IEEEtitlepagestyle%
 \def\@oddhead{\strut\hfill#1\hfill\strut}%
 \def\@evenhead{\strut\hfill#1\hfill\strut}%
 }%
 \ps@headings%
 }
 \makeatother

\confheader{%
 2024 International Conference on Integrated Circuits and Communication Systems (ICICACS)
 }

 \usepackage[pscoord]{eso-pic}
\newcommand{\placetextbox}[3]{
 \setbox0=\hbox{#3}
 \AddToShipoutPictureFG*{ \put(\LenToUnit{#1\paperwidth},\LenToUnit{#2\paperheight}){\vtop{{\null}\makebox[0pt][c]{#3}}}
 }
 }
 \placetextbox{.23}{0.055}{\small{979-8-3503-1755-8/24/\$31.00~\copyright 2024 IEEE}}

\usepackage{cite}
\usepackage{amsmath,amssymb,amsfonts}
\usepackage{algorithmic}
\usepackage{graphicx}
\usepackage[colorlinks=true, allcolors=blue]{hyperref}
\usepackage{textcomp}
\def\BibTeX{{\rm B\kern-.05em{\sc i\kern-.025em b}\kern-.08em
    T\kern-.1667em\lower.7ex\hbox{E}\kern-.125emX}}
\begin{document}

\title{Analyzing Musical Characteristics of National Anthems in Relation to Global Indices\\}

\author{\IEEEauthorblockN{S M Rakib Hasan}
\IEEEauthorblockA{\textit{Dept. of CSE} \\
\textit{BRAC University}\\
Dhaka, Bangladesh \\
sm.rakib.hasan@g.bracu.ac.bd}
\and

\IEEEauthorblockN{Aakar Dhakal}
\IEEEauthorblockA{\textit{Dept. of CSE} \\
\textit{BRAC University}\\
Dhaka, Bangladesh \\
aakar.dhakal@g.bracu.ac.bd}
\and
\IEEEauthorblockN{Ms. Ayesha Siddiqua}
\IEEEauthorblockA{\textit{Dept. of ECE} \\
\textit{Nitte Meenakshi IT}\\
Bengaluru, India\\
ayesha.siddiqua@nmit.ac.in}
\and

\IEEEauthorblockN{Mohammad Mominur Rahman}
\IEEEauthorblockA{\textit{Division of ICT}\\
{College of S and E}\\
\textit{Hamad Bin Khalifa University}\\
Doha, Qatar \\
mora28982@hbku.edu.qa}
\and

\IEEEauthorblockN{Md Maidul Islam}
\IEEEauthorblockA{\textit{
Department of ECE}\\
\textit{Green University of Bangladesh}\\
Dhaka, Bangladesh \\
engr.maidul.eee@gmail.com}
\and

\IEEEauthorblockN{Mohammed Arfat Raihan Chowdhury}
\IEEEauthorblockA{\textit{College of S and E} \\
\textit{Hamad Bin Khalifa University}\\
Doha, Qatar \\
arfatmarine27@gmail.com}
\and 

\IEEEauthorblockN{ S M Masfequier Rahman Swapno}
\IEEEauthorblockA{\textit{Dept. of CSE} \\
\textit{BUBT}\\
Dhaka, Bangladesh \\
masfequier.cse.bubt@gmail.com}
\and
\IEEEauthorblockN{SM Nuruzzaman Nobel}
\IEEEauthorblockA{\textit{Dept. of CSE} \\
\textit{BUBT}\\
Dhaka, Bangladesh\\
smnuruzzaman712@gmail.com}}

\maketitle

\begin{abstract}
 Music plays a huge part in shaping peoples' psychology and behavioral patterns. This paper investigates the connection between national anthems and different global indices with computational music analysis and statistical correlation analysis. We analyze national anthem musical data to determine whether certain musical characteristics are associated with peace, happiness, suicide rate, crime rate, etc. To achieve this, we collect national anthems from 169 countries and use computational music analysis techniques to extract pitch, tempo, beat, and other pertinent audio features. We then compare these musical characteristics with data on different global indices to ascertain whether a significant correlation exists. Our findings indicate that there may be a correlation between the musical characteristics of national anthems and the indices we investigated. The implications of our findings for music psychology and policymakers interested in promoting social well-being are discussed. This paper emphasizes the potential of musical data analysis in social research and offers a novel perspective on the relationship between music and social indices. The source code and data are made open-access for reproducibility and future research endeavors. It can be accessed at \href{http://bit.ly/na_code}{http://bit.ly/na\_code}
\end{abstract}


\section{Introduction}
A country's national anthem is a symbolic representation of its identity, culture, and values. It is a piece of music that is intended to inspire patriotism, elicit emotion, and unite people under a common identity. However, the effect of national anthems on society is poorly understood, particularly in relation to larger societal issues such as peace and mental health. The purpose of this paper is to investigate the possible relationship between national anthems and societal indices.\\
National anthems, like music in general, have long been acknowledged as a potent influence on attitudes and behaviors. National anthem melodies, patterns, and lyrics can elicit powerful emotional responses and reinforce cultural norms and values. People listen to the national anthems from their childhood more than any other song, which can permanently shape their psychological characteristics. It is unclear, however, to what extent national anthems can affect societal outcomes. By analyzing the musical characteristics of national anthems and comparing them to statistics on world peace and suicide rates, we expect to cast light on the potential impact of national anthems on these significant societal issues.\\
To achieve this, we will employ computational music analysis techniques to extract pertinent characteristics from national anthems, such as intonation, cadence, and other musical characteristics. Then, we will compare these characteristics with data on the world peace index, world crime rate index, Human Development Index (HDI), Happiness index and the World suicidal index to determine if there is a significant correlation between the musical characteristics of national anthems and these societal indices.\\
This investigation has significant implications for policymakers and musicians equally. If we can identify the musical characteristics that are linked to positive social outcomes such as harmony and mental health, we can use this information to promote positive change through music. By better comprehending the role national anthems play in influencing societal attitudes and behaviors, we can work towards a more peaceful and harmonious global community. With musical analysis, the generative audio industry can also produce personalized musical pieces.\\
In this paper, we discuss previous works in this field, the characteristics of musical data, our methodology, experimental setup, and results.

\section{Previous Work}
For the song analysis, Machine learning (ML) has been used largely through ML used in AI healthcare \cite{b28}, \cite{b27} \cite{17} \cite{b23} \cite{b22} business \cite{b19},\cite{b20}, \cite{b21} and many more \cite{b25}, \cite{b26}, \cite{b29}. However, in our work, we choose a statistical approach.
The research\cite{b1} was conducted on over one million users with more than 200,000 songs. They used the million-song dataset\cite{b2} to analyze the fundamental structure of millions of users' musical preferences by analyzing massive amounts of data from natural music listening environments.  Arousal, valence, and depth (AVD) are the three broad dimensions that underpin people's musical feature preferences, as confirmed by large-scale data and prior theory. The arousal component defines a preference for high-energy music, that is, boisterous, rousing music that does not include acoustic music. The valence component identifies songs that are perceived as being enjoyable, upbeat, and frequently danceable. The depth dimension has diminished clarity.\\

A previous study\cite{b3} examines the structure of Dutch adolescents' music preferences, the stability of music preferences, and the relationships between the Big-Five personality traits (Extraversion, Agreeableness, Conscientiousness, Emotional stability, and openness) and changes in music preferences. About one thousand students between the ages of 12 and 19 were asked by hand to complete a questionnaire, which revealed four genres that were easily interpretable: rock, elite, urban, and pop/dance. In addition to being comparatively stable over 1-, 2-, and 3-year intervals, music preferences were consistently associated with personality traits, corroborating previous research conducted in the United States. Furthermore, personality traits were found to predict variations in music preferences over three years.\\

Using the Five-Factor Model of personality, the paper\cite{b4} explores the relationship between musical preference and personality. According to this paradigm, personality can be measured along five dimensions: openness, conscientiousness, extraversion, agreeableness, and neuroticism. In a study conducted by the authors, participants were asked to complete a personality questionnaire and then select their musical preferences from a selection of 18 musical genres. Their findings revealed a relationship between musical preference and personality, with certain musical genres associated with particular personality attributes. Those who scored highly on openness tended to favor jazz, classical, and folk music, whereas those who scored highly on extraversion tended to prefer rock, rap, and dance music. In addition, the authors discovered that heavy metal and country music were less significantly associated with any specific personality trait.\\

These findings are consistent with the study\cite{b5}, which also used the Five-Factor Model of personality to investigate the relationship between musical preference and personality, these findings demonstrate a correlation between musical preference and personality. In this study, they used a larger sample size and a wider variety of musical genres. They discovered a correlation between musical preference and personality, with certain musical genres associated with particular personality traits. Those who scored highly on openness tended to favor blues, jazz, and classical music, whereas those who scored highly on extraversion tended to prefer rock, rap, and heavy metal. The disparities between studies' findings suggest that the relationship between musical preference and personality may be complex and nuanced; therefore, additional research is required to fully comprehend this relationship.\\

The purpose of the study\cite{b6} was to examine the relationship between personality traits and musical genre preferences among Japanese college students. 268 students, 153 males, and 115 females, submitted a questionnaire assessing their personality traits and general music genre preferences. Using the HEXACO-100, which measures six personality domains, each of which encompasses four more specific aspects, the personality traits were evaluated. Jazz, classical, opera, gospel, enka, pop, rock, punk, hip-hop, R\&B, techno, and reggae were evaluated as musical genres. Openness and aesthetic appreciation were associated with a preference for reflective music genres, whereas sociability was linked to a preference for contemporary music. Other personality dimensions were less consistently associated with musical preferences, indicating the need for more specific assessments of both personality and music genres.\\

Using both self-report measures and Facebook likes, the study\cite{b7} sought to investigate the relationship between musical preferences and personality characteristics. Two investigations were conducted by the authors, one using a self-report questionnaire and the other using Facebook likes. In the first study, participants filled out a questionnaire that assessed their Big Five personality traits and their musical preferences across 26 genres. The results indicated that each genre was associated with a distinct personality trait pattern. Those who preferred reflective and complex music (such as jazz, classical, and folk) scored higher on openness to experience, whereas those who preferred upbeat and conventional music (such as country and pop) scored higher on extraversion and lower on openness to experience. In the second study, the authors analyzed the Facebook likes of over 79,000 people to predict their personality traits and musical tastes. The results demonstrated that Facebook likes accurately predicted both personality traits and musical preferences. In addition, the analysis revealed that musical preferences were a more accurate predictor of personality than demographic variables like age and gender.\\

The study\cite{b8} examined the connection between musical preferences and personality. They asked over 500 participants to rate their level of delight for 120 musical themes and genres and also administered the Sixteen Personality Factor Questionnaire (16PF). There were statistically significant correlations between certain musical preferences and personality characteristics, according to the findings. Those who preferred classical and jazz music, for instance, scored higher on measures of intelligence, openness to experience, and emotional stability, whereas those who preferred country and Western music scored higher on measures of extraversion and lower on measures of intelligence. Gender and age also affected musical preferences, with women showing a stronger preference for classical music than males and older individuals demonstrating a stronger preference for music from the "Golden Age." Even though the study has limitations, such as a lack of diversity in the participant pool, it provides evidence for the connection between musical preferences and personality characteristics, which has implications for music therapy and other interventions that use music as a tool.\\

The research\cite{b9} sought to validate the German version of the Short Test of Music Preferences (STOMP) and investigate the relationship between music preferences and personality. Strong correlations were found between the STOMP factors and their corresponding audio files, proving the instrument's validity. The study also revealed a consistent pattern of associations between musical preferences and personality, with Openness and Extraversion serving as the most accurate predictors of musical preferences. Specifically, individuals who were more open to new experiences tended to favor Reflective \& Complex, Intense \& Rebellious music, while extraverted individuals preferred Upbeat \& Conventional, Energetic \& Rhythmic, and Intense \& Rebellious music. The results also indicated that neuroticism was negatively associated with Intense and Rebellious music, that agreeableness only affected ratings of Upbeat and Conventional music, and that conscientiousness had no effect on music preferences. Overall, the research contributes to the body of knowledge by validating the STOMP in a German sample and replicating previous findings regarding the association between musical preferences and personality.\\

The paper\cite{b10} examines music preference and adolescent mental health. Music helps adolescents cope with identity, values, and self-perception by expressing or validating their feelings. Heavy metal music has been linked to assertiveness, aggression, and impulsivity in several studies, which the paper reviews. The paper also discusses studies that disagree that music causes suicide but suggest that certain music preferences may increase suicide risk. In girls, rock/metal music preferences were linked to suicidal thoughts, depression, delinquency, drug use, and family dysfunction. The music primed implicit suicide cognitions without increasing suicide risk. Non-violent rap songs caused more depression than violent ones.\\

The research\cite{b11} investigated the correlation between the emotional valence of national anthems and suicide rates. A group of 30 American students were presented with the national anthems of 18 European nations and were asked to rate each anthem based on its level of melancholy and sadness. Additionally, the proportion of low notes in each anthem was computed. The present study examined the correlation between scores and suicide rates for the year 2000, disaggregated by gender, as reported by the World Health Organization. The findings of the study indicated a positive correlation between the prevalence of low-pitched notes in national anthems and elevated suicide rates, thereby lending support to Rihmer's proposal. Nevertheless, this study is subject to certain limitations, including the impact of extraneous variables on suicide rates and the potential for evaluators' bias.\\

The objective of the study\cite{b12} was to examine the correlation between personality traits and music preferences through the direct measurement of music listening behavior. The study involved a sample of 395 individuals who were employed by Royal Philips Electronics, and whose ages ranged from 22 to 60 years. The study involved administering a survey to the participants to gather demographic information, personality traits, and music preferences. Additionally, the music-listening behavior of the participants was monitored for a minimum of three months using a music database. The findings of the study revealed a favorable association between self-reported music preferences and listening behavior, thereby suggesting a connection between personality and music preference. Furthermore, the research aimed to validate Rentfrow and Gosling's music preference model, however, the framework was not corroborated. The research findings indicate that the utilization of genre labels as a metric for gauging music preferences is problematic. The study recommends that further investigation be conducted to delve deeper into these issues.

In the study \cite{b16} the focus is on advancing Music Information Retrieval (MIR), specifically in the challenging domain of chord recognition for modern rock and metal music. Addressing the limitations of existing methods, the paper introduces an innovative approach using a neural network and the Pitch Class Profile vector for more efficient chord recognition. This vector, comprising 12 semi-tone values, simplifies the process amidst the complexity of the music genres in question. Additionally, the research employs the REPET (REpeating Pattern Extraction Technique) method for distinct audio separation, enhancing chord identification from monophonic and polyphonic audio. The paper's methodology, emphasizing the Constant Q Transform for feature extraction and template matching for chord correlation, marks a significant stride in the field of automatic chord recognition, particularly in processing intricate music styles.

The paper \cite{b16} \cite{17} delves into the evolving field of Music Information Processing, highlighting the importance of multimodal approaches that integrate diverse data types like audio recordings, music scores, motion data, and cultural tags. It traces the historical progression of music representation systems, underscoring the influence of technological advancements on the development of sophisticated music description methods. The study categorizes various multimodal tasks, emphasizing the breadth of applications from audio-to-score alignment to emotion recognition. Finally, it identifies key challenges facing the field, such as the scarcity of comprehensive multimodal music datasets and the need for advanced algorithms to align and represent the rich multimodality of music.

\section{Methodology}

\subsection{Data Collection}
In this study, we have collected two distinct categories of data. The initial category is national anthems. We compiled as midi files the national anthems of 171 countries from Kaggle. Our subsequent set of data included the World Peace Index, the World Suicide Rate Index, the World Crime Index, the World Happiness Index, and the World Human Development Index. All of the information was obtained from the website www.worldpopulationreview.com.

\subsection{Data Preprocessing}
Our midi dataset was preprocessed so that any empty files were removed. Additionally, multiple files existed for the same country. After preprocessing, we were left with midi files containing the national anthems of 166 countries. The index data contained a great deal of information deemed superfluous for our investigation. After removing these, we had the country index scores and ranks. The number of nations varied according to the index. Finally, we eliminate the countries that are not present in both datasets and retain the data for the countries that are unique to each dataset.

\subsection{Computational Music Analysis}
We chose a few musical characteristics that distinguish one song from another. Utilizing Python libraries Music21 \cite{b13} and pretty-midi, the features were extracted. 
\newline Music21 is a Python library for music theory and analysis. It provides a wide range of features for extracting musical features, such as pitch, rhythm, and harmony. It also allows to the creation of data frames and databases of musical data and visualization of musical characteristics. We have used this to extract harmonic, melodic, and rhythmic analysis of the national anthems by converting them to midi stream objects. The numerical notations of the musical features allowed us to computationally analyze the anthems. Also, it is a great tool for visualizing musical characteristics. We can directly distinguish among the music by just looking at the plots generated.
\newline PrettyMIDI is another tool that we have used to collect numerical data about the instrument, tempo, and dynamics of the music. This provides us insights into the overall expressive qualities of the anthems.

Our extracted chosen features are described as such:

\begin{itemize}

\item Melodic Contour: This refers to the structure or pattern of a melody over time, specifically how its pitch raises and declines. It is one of the primary ways a melody creates a sense of tension and release, and it can convey emotions such as happiness, sorrow, and exhilaration.

\item Pitch: The pitch of a sound is its perceived highness or lowness. The frequency of sound waves determines the pitch in music, with higher frequencies corresponding to higher pitches and lower frequencies corresponding to lower pitches. Pitch is an essential element of melody and harmony, and it can be used to generate contrast, tension, and resolution in musical compositions.

\item Beat: Beat refers to the underlying pulse of a musical composition, which is often perceived as a steady, repetitive pattern of accents or stresses. The rhythm provides a basis for the melody and harmony and contributes to the overall mood and groove of the music. Rhythmic patterns and tempos can vary significantly between musical genres, contributing to their distinctive sound and manner.

\begin{figure}[h!]
\centerline{\includegraphics[width=0.5\textwidth]{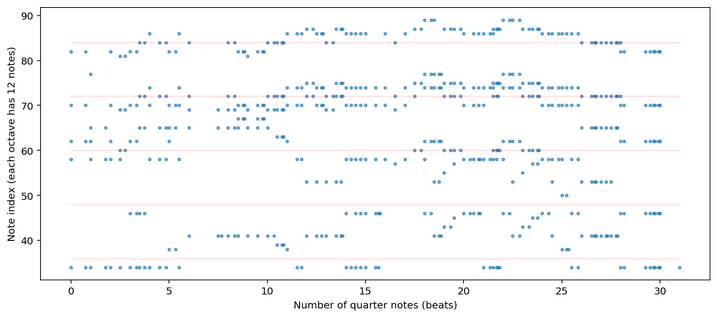}}
\caption{Octave and Beat distribution of the most suicide-prone country(Lesotho)}
\label{fig2}
\end{figure}

\begin{figure}[h!]
\centerline{\includegraphics[width=0.5\textwidth]{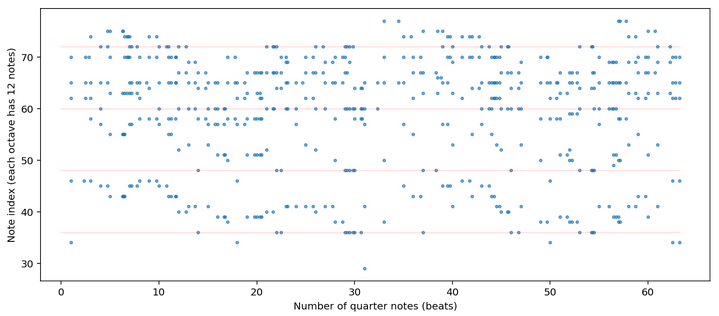}}
\caption{Octave and Beat distribution of the least suicide-prone country(Antigua and Barbuda)}
\label{fig2}
\end{figure}

\item Tempo: It is the speed or cadence at which a piece of music is played, and is typically measured in beats per minute (BPM). Tempo can have a significant effect on the mood and energy of a piece of music, with slower tempos often generating a more relaxed or contemplative atmosphere and faster tempos a more energizing or upbeat atmosphere. Different musical genres are frequently associated with distinct tempo ranges.

\item Note Velocity: Note velocity, also referred to as note intensity, is the force with which a note is performed. Note velocity is typically conveyed as a value between 0 and 127 in MIDI sequencing, with greater values indicating louder and more forceful performance.

\item Note Duration: The duration of a note refers to how long a note is held or sustained. Typically, in music notation, note duration is denoted by the shape of the note symbol (e.g., whole note, half note, quarter note), with lengthier notes having more "filled-in" shapes. Variable note duration can be used to generate rhythmic patterns and syncopation and to add complexity and intrigue to a composition.

\item Rest Duration: Rest duration refers to the duration of a rest symbol (silent) in a composition. In music notation, the duration of rest is typically represented by the shape of the rest symbol (e.g., whole rest, half rest, quarter rest), with lengthier rests having more "filled-in" shapes. Rest duration can be varied to create rhythmic interest and tension, as well as pauses and interruptions in a composition.

\item Time Signature Change: Time signature changes indicate meter changes by defining the number of beats per measure and their length. To produce contrast, variety, tension, or relaxation, time signatures might change inside measures, between measures, or throughout a piece. Changing time signatures complicates a piece yet makes it harder to follow. 

\end{itemize}

In order to gain a comprehensive comprehension of the midi file's characteristics, we considered the mean of note duration, melodic contour, and beats, the median of note velocity and rest durations, mode of pitch along with the estimated tempo and the time signature changes.

\subsection{Clustering}
Using the elbow and silhouette methods to determine the optimal number of clusters, we cluster both the national anthems and the index data based on the characteristics and information of the datasets. Before the clustering, the rankings and country names were eliminated. The anthems and indices were grouped using the Kmeans clustering algorithm.

\subsection{Correlation Analysis}
We carried out a correlation study by contrasting the country categories presented in the national anthems with those presented in the index data. The technique is repeated for each index, and the projected correlation with each musical quality is computed after each iteration. To visually demonstrate the association, we use heat maps.\\

\begin{figure}[h!]
\centerline{\includegraphics[width=0.5\textwidth]{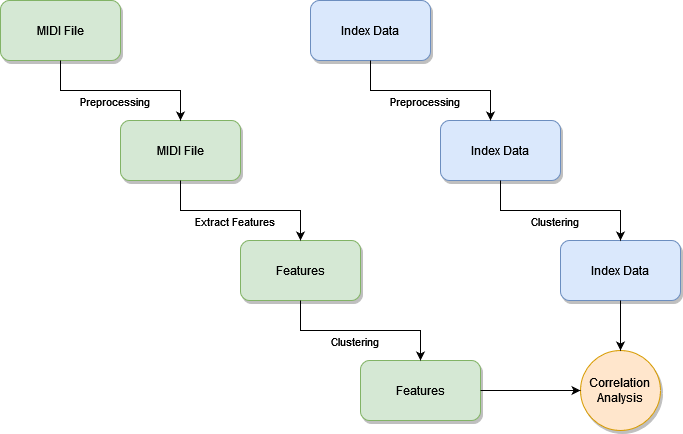}}
\caption{Research Workflow}
\label{fig1}
\end{figure}

Our workflow is shown in figure \ref{fig1}

\section{Results}

Based on the correlation analysis, we've determined that there are significant correlations between musical characteristics and certain global indices. Unfortunately, there are not many studies available to compare our performance with. However, we believe that future scopes can pave the way to formulate a baseline for this type of study. Our results, for each of the indices, are tabulated below:

\begin{table}[htbp]
\caption{Suicide Index}
\begin{center}
\begin{tabular}{|c|c|c|c|c|c|}
\hline
\textbf{}&\multicolumn{2}{|c|}{\textbf{Suicide Rates}} \\
\cline{2-3} 
\textbf{Musical Characteristics} & \textbf{\textit{Low}}& \textbf{\textit{High}} \\
\hline
        Melodic Contour & ~Low & High~ \\ \hline
        Pitch & ~Low & High~ \\ \hline
        Beat & ~High & Low~ \\ \hline
        Tempo &  ~High & Very Low~ \\ \hline
        Note Velocity &  ~High & Low~ \\ \hline
        Note Duration & ~Low & High~ \\ \hline
        Rest Duration & ~High & Low~ \\ \hline
        Time Signature Changes & ~Average & More \\ \hline
\end{tabular}
\label{tab1}
\end{center}
\end{table}
\begin{table}[htbp]
\caption{Happiness Index}
\begin{center}
\begin{tabular}{|c|c|c|c|c|c|}
\hline
\textbf{}&\multicolumn{2}{|c|}{\textbf{Happiness Score}} \\
\cline{2-3} 
\textbf{Musical Characteristics} & \textbf{\textit{Low}}& \textbf{\textit{High}} \\
\hline
Melodic Contour & Low  & High \\ \hline
        Pitch & High & Low \\ \hline
        Beat & Low  & Very High \\ \hline
        Tempo & Low  & Very High \\ \hline
        Note Velocity & High & Low \\ \hline
        Note Duration & High & Low \\ \hline
        Rest Duration & Low & High \\ \hline
        Time Signature Changes & High & Low \\ \hline
\end{tabular}
\label{tab1}
\end{center}
\end{table}
\begin{table}[htbp]
\caption{Peace Index}
\begin{center}
\begin{tabular}{|c|c|c|c|c|c|}
\hline
\textbf{}&\multicolumn{2}{|c|}{\textbf{Peace Score}} \\
\cline{2-3} 
\textbf{Musical Characteristics} & \textbf{\textit{Low}}& \textbf{\textit{High}} \\
\hline
Melodic Contour & Low  & High \\ \hline
        Pitch & Slightly High & Slightly Low \\ \hline
        Beat & Low  & High \\ \hline
        Tempo & High & Low \\ \hline
        Note Velocity & High & Low \\ \hline
        Note Duration & Low & High \\ \hline
        Rest Duration & High & Low \\ \hline
        Time Signature Changes & High & Low \\ \hline
\end{tabular}
\label{tab1}
\end{center}
\end{table}
\begin{table}[htbp]
\caption{Crime Index}
\begin{center}
\begin{tabular}{|c|c|c|c|c|c|c|}
\hline
\textbf{}&\multicolumn{2}{|c|}{\textbf{Crime Rate}} \\
\cline{2-3} 
\textbf{Musical Characteristics} & \textbf{\textit{Low}}& \textbf{\textit{High}} \\
\hline
Melodic Contour & Very High & Low \\ \hline
        Pitch & Slightly Low & Slightly High \\ \hline
        Beat & Low & High \\ \hline
        Tempo & Low & High \\ \hline
        Note Velocity & Very High & Low \\ \hline
        Note Duration & High & Low \\ \hline
        Rest Duration & Low & High \\ \hline
        Time Signature Changes & More & Average \\ \hline
\end{tabular}
\label{tab1}
\end{center}
\end{table}
\begin{table}[htbp]
\caption{Human Development Index}
\begin{center}
\begin{tabular}{|c|c|c|c|c|c|c|}
\hline
\textbf{}&\multicolumn{2}{|c|}{\textbf{Human Development Score}} \\
\cline{2-3} 
\textbf{Musical Characteristics} & \textbf{\textit{Low}}& \textbf{\textit{High}} \\
\hline
Melodic Contour & Slightly High & Average \\ \hline
        Pitch & Slightly Low & Average \\ \hline
        Beat & Slightly Low & High \\ \hline
        Tempo & Average & Average \\ \hline
        Note Velocity & High & Average \\ \hline
        Note Duration & Average & High \\ \hline
        Rest Duration & Low  & Very Low \\ \hline
        Time Signature Changes & Average & Average \\ \hline
\end{tabular}
\label{tab1}
\end{center}
\end{table}
The findings indicate a significant association between pitch, beat, tempo, note duration, and rest duration with suicide rate, peace score, and happiness score. There appears to be a negligible association between the national anthems and the crime rate as well as the human development index. Research suggests that certain factors, such as low pitch, high tempo, high beat, low note duration, and high rest duration, may be associated with lower suicide rates and higher scores for happiness and peace. The correlation of the Happiness Index with the National Anthems is shown in figure \ref{fig2}.

\begin{figure}[h!]
\centerline{\includegraphics[width=0.5\textwidth]{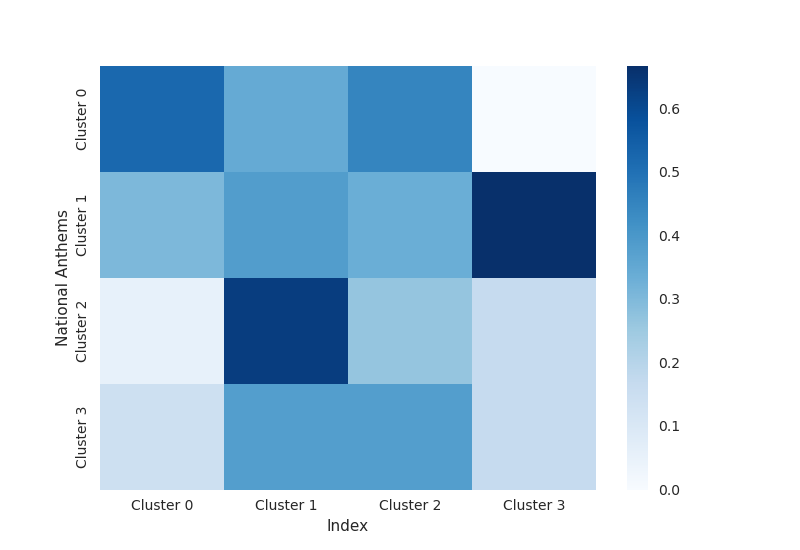}}
\caption{National Anthem and Happiness Index}
\label{fig2}
\end{figure}

\section{Discussion and Future Scopes}
Compared to some previous studies discussing musical preferences and human personalities, this is a completely different approach to musical analysis. All the previous studies concerning music and human psychology have taken primary data by conducting surveys. Moreover, these are done as individual personality analyses. They have taken the musical preferences of people, their personality traits, and their age. All these studies are done in the Western World. However, the possibility of imputing wrong or biased information is also not ruled out. In this research, however, we have discarded human interaction and completely based our process on computation and numerical data directly parsed from the music. We tried to predict the Also, this study does not involve any learning process that may cause some bias. Therefore it stands robust unlike most of the previous studies. However, the odds of confirmation bias\cite{b14} and Texas Sharpshooter Fallacy\cite{b15} are not completely ruled out. 
\newline This research only investigates the chances of correlation, not causation. Though we have found a very significant correlation in some aspects, we cannot conclude that national anthems are the cause behind the countries’ global indices. The national anthem does not represent a country’s entire culture rather only provides a slight insight into it. We have decided to take the national anthem because it is the most sung and heard song of every country, so it must have a deep imprint in the minds of its people. 
\newline Future studies in this field should include more musical samples, preferably the most popular ones from each country, and analyze them to conclude the effect of music on an entire country. The musical analysis can be done on a personal level too. This can determine the psychological traits of a person or a group which can be useful in different aspects.

\section{Conclusion}
In conclusion, our study provides evidence of a potential relationship between the aural characteristics of national anthems and a variety of social indices, such as the rate of suicide, the level of happiness, and peace. Certain musical characteristics, such as low pitch, high tempo, and high beat, may be associated with positive social outcomes, according to the findings. The implications of these findings for policymakers and researchers interested in promoting social well-being and music psychology are substantial. In addition, this study emphasizes the potential of employing computational music analysis techniques in social research, offering a fresh perspective on the relationship between music and social indices. To confirm these correlations and investigate the underlying mechanisms of the relationship between music and social well-being, however, additional research is required. Future research opportunities in this field may include investigating additional social outcomes, inspecting additional musical characteristics, and analyzing additional musical genres.
\section{Acknowledgement}
The authors would like to express their gratitude to Sakib Ul Rahman Sourove, Department of Computer Science and Engineering, Dhaka University, for his advice and support. His insights and guidance have been instrumental in shaping the direction of this research. Also, they would like to thank Shah Sakib Sadman Pranto, from the Department of Public Administration, Dhaka University, whose expertise as a musician has been invaluable in helping determine the musical features to select for this research.

\end{document}